\documentclass[%
 reprint,
 amsmath,amssymb,
 aps,
]{revtex4-2}

\usepackage{graphicx}% Include figure files
\usepackage{dcolumn}% Align table columns on decimal point
\usepackage{bm}% bold math
\usepackage{mathtools} 
\usepackage{amsfonts}
\usepackage{amsmath}
\usepackage{graphicx}
\usepackage{array}
\usepackage{multirow}
\usepackage{verbatim}
\newcommand{\ud}{{\rm d}}

\newcommand{\ii}{{\rm i}}
\DeclareMathOperator{\Tr}{Tr}
\newcommand{\be}[0]{\begin{equation}}
\newcommand{\ee}[0]{\end{equation}}
\newcommand{\bea}[0]{\begin{eqnarray}}
\newcommand{\eea}[0]{\end{eqnarray}}

\begin{document}

\preprint{APS/123-QED}

\title{Coherence effects on estimating two-point separation}% Force line breaks with \\
%\thanks{A footnote to the article title}%

\author{Kevin Liang}
% \altaffiliation[Also at ]{Physics Department, XYZ University.}%Lines break automatically or can be forced with \\
 \email{kliang3@ur.rochester.edu}
\author{S. A. Wadood}%

\affiliation{The Institute of Optics and Center for Coherence and Quantum Optics, University of Rochester, 275 Hutchinson Rd, Rochester, NY 14627, USA
}%

%\affiliation{Center for Coherence and Quantum Optics, University of Rochester, 275 Hutchinson Rd, Rochester, NY 14627, USA}

\author{A. N. Vamivakas}
\affiliation{The Institute of Optics, and Center for Coherence and Quantum Optics, and Department of Physics and Astronomy, and Materials Science, University of Rochester, 275 Hutchinson Rd, Rochester, NY 14627, USA
}%

%\affiliation{Center for Coherence and Quantum Optics, University of Rochester, 275 Hutchinson Rd, Rochester, NY 14627, USA}
%\affiliation{Department of Physics and Astronomy, University of Rochester, 275 Hutchinson Rd, Rochester, NY 14627, USA}
%\affiliation{Materials Science, University of Rochester, 275 Hutchinson Rd, Rochester, NY 14627, USA}

%\collaboration{MUSO Collaboration}%\noaffiliation

\date{\today}% It is always \today, today,
             %  but any date may be explicitly specified

\begin{abstract}
The quantum Fisher information (FI), when applied to the estimation of the separation of two point sources, has been shown to be non-zero in cases where the coherence between the sources are known. Although it has been claimed that ignorance of the coherence causes the quantum FI to vanish (a resurgence of Rayleigh's curse), a more complete analysis including both the magnitude and phase of the coherence parameter is given here. Partial ignorance of the coherence is shown to potentially break Rayleigh's curse, whereas complete ignorance guarantees its resurgence.
\end{abstract}

%\keywords{Suggested keywords}%Use showkeys class option if keyword
                              %display desired
\maketitle

%\tableofcontents

\section{Introduction}

It is intuitive that, in the problem of estimating the separation of two point sources, the closer the sources become, the more difficult the task \cite{Rayleigh1879}. This limitation is intimately related to the study of resolution, whose definition has been the topic of discussion in recent investigations. Among the earliest definitions was the notion of the minimum separation between two point sources for which they can be distinguished with an imaging system. In this classical context, resolution was directly related to the maximum spatial frequency content that is adequately passed by the imaging system (which in turn is related to the system's aperture) \cite{Goodman,denDekker1997}. The term \textit{superresolution} in this context, which is an ongoing area of research, refers to concepts that allow for the production of images where two point sources are resolved to be closer than is allowed by the classical resolution limit; various techniques include, but are not limited to, stochastic optical reconstruction microscopy (STORM) \cite{hell2007review}, stimulated emission depletion microscopy (STED) \cite{hell2007review}, and using superoscillations in conjunction with confocal microscopes \cite{GburReview,Smith2016, Aharonov1988,Kosmeier2011}. However, in the context of image processing, resolution takes the definition of the minimum separation between point sources that can be estimated with a desired precision \cite{denDekker1997}. It is this latter definition of resolution that is the focus of this manuscript, which uses the language of the Fisher information (FI) \cite{Helmstrom}. The \textit{classical} FI quantifies the upper-bound of information that one can obtain regarding the separation of the two point sources, for a specified measurement scheme \cite{Kay}. For the simplest schemes (such as the direct imaging of two point sources through an optical system), the classical FI vanishes as the separation vanishes; this phenomenon has been termed Rayleigh's curse.

However, it has recently been shown that it is possible to circumvent Rayleigh's curse through the implementation of new measurement schemes. Such developments were based on the \textit{quantum} FI \cite{Helmstrom}, which provides an experiment-free information lowerbound on the two-point separation problem. Quantum FI calculations showed that it was theoretically possible to obtain a non-vanishing information lower-bound even as the source separation approaches zero \cite{Tsang2016}. The proposal in Ref.~\onlinecite{Tsang2016} for incoherent point sources was verified in a series of experiments \cite{OptExpress_Parity_Sorting2016,sanchezsoto_2016_optica,superresolution_with_heterodyne,steinberg2017beating_rayleighscurse}. These results were followed by other experiments and discussions that commented on the true attainability of the quantum FI (the reciprocal of the variance upperbound of a parameter estimation) \cite{Larson2018,Wadood2020}. In particular, the coherence between the two point sources was identified as a relevant parameter to explore.

When the coherence, along with the separation, between two point sources is unknown, one must account for the need to jointly estimate both the coherence and the separation in a multi-parameter estimation framework. Previous work claimed, through quantum FI calculations, that the ignorance of coherence gives rise to the return of Rayleigh's curse \cite{Larson2018}. However, the theoretical framework of the analysis was valid only for real coherence parameters. Because of this, the phase of the coherence was implicitly assumed to be known. In the language of multi-parameter estimation, the coherence between two point sources is complex and therefore is comprised of two real parameters. Previous works therefore implicitly assumed that one of these parameters (the phase) is known, an assumption that we relax in our work.

A complete analysis of the effects of coherence in estimating the separation of two point sources is provided in this work. By allowing the coherence to be expressed as two real parameters (either or both of which may be unknown), novel phenomena, regarding the behavior of the quantum FI when the separation of the two point sources vanishes, are revealed. The discussion regarding whether Rayleigh's curse persists when coherence is unknown is shown to be more complicated than initially expected. In addition to clarifying the effects of coherence on two-point separation estimation, the results presented here also serve to append the known results in the literature regarding the quantum FI of separation estimation for arbitrary coherence. So far, the cases where the sources are incoherent, real and partially coherent, and fully coherent have been studied \cite{Tsang2016,Larson2018,Hradil2019}. This work provides the framework necessary to fill out the remainder of the complex coherence disk as shown in Fig. \ref{introFig}(b). In doing so, as this work's results demonstrate, we obtain both a deeper understanding of the connections of the previously studied cases and novel phenomenon only evident in a full treatment of the complex coherence. For example, we find a singular/multivalued behavior in the quantum FI for some scenarios of multiparameter estimation.

It is important to point out that the quantum FI framework has been used extensively to study other generalizations of the incoherent two-point separation problem \cite{Jing2019,Zhou2019}. These include extending the incoherent treatment to more than two point sources and finding information bounds for separation estimation in three-dimensional space. Other parameters that have been analyzed are the estimation of longitudinal separation \cite{yiyu2019axial_superresolution_optica}, simultaneous estimation of the separation, centroid, and relative intensities between two incoherent point sources \cite{Rehacek2017}.

\section{Theory}
\label{theory}
\begin{figure}
    \centering
    \includegraphics[scale=0.5]{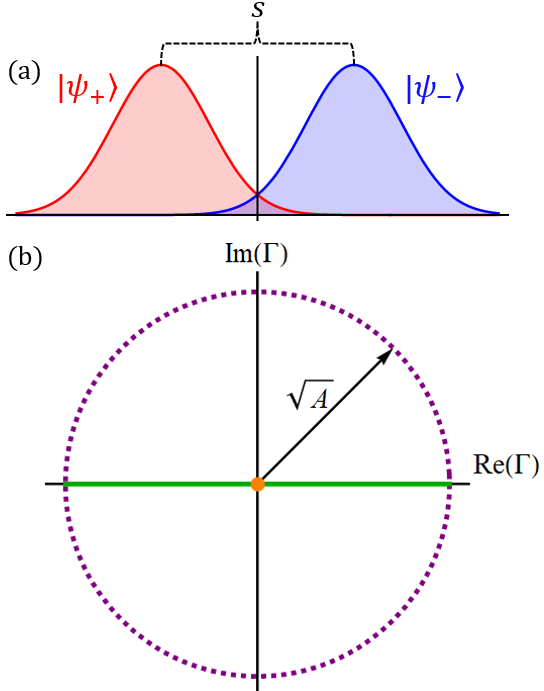}
    \caption{Two Gaussian PSFs separated by $s$, corresponding to the kets $|\psi_\pm \rangle$, are shown in (a). The complex $\Gamma = r e^{\ii \phi}$ disk is shown in (b), where the possible values of $\Gamma$ is bounded by the condition $|\Gamma| \le \sqrt{A}$. The specific cases of $\Gamma =0$ (orange point), $\text{Im}(\Gamma) = 0$ (green line) and $|\Gamma| = \sqrt{A}$ (dashed purple circle) are among the cases previously analyzed. }
    \label{introFig}
\end{figure}

Our treatment begins in a fashion that is analogous to the approaches of \cite{Tsang2016,Larson2018,Hradil2019} by stipulating an image-plane density matrix in the not-orthogonal (as indicated by the subscript) basis $\{|\psi_+ \rangle, |\psi_- \rangle\}$:
\begin{align}
    \hat{\rho} = \frac{1}{(1+A)[1+d^2(s)] + 4 d(s) \text{Re}(\Gamma)}\begin{bmatrix} 1 & \Gamma  \\ \Gamma^* & A \end{bmatrix}_{\text{no}}, \label{densMat}
\end{align}
where $A$ is the intensity ratio of the two point sources, and the (unnormalized) coherence parameter is given by $\Gamma \triangleq r e^{\ii \phi}$ with $r \le \sqrt{A}$. A derivation of Eq.~(\ref{densMat}) is given in Appendix~\ref{app1}. The basis kets are defined over position states as
\begin{align}
    |\psi_\pm \rangle &\triangleq \int_{-\infty}^{\infty} \ud x \, \psi\left( x \pm \frac{s}{2} \right) |x \rangle,
\end{align}
where $s$ is the separation between the two point sources, as seen in Fig.~\ref{introFig}(a), and $\psi$ represents the point spread function (PSF) of the imaging system used to resolve the two point sources. As mentioned earlier, $|\psi_+\rangle$ and $|\psi_-\rangle$ are not orthogonal since
\begin{align}
    d(s) &\triangleq \langle \psi_+|\psi_- \rangle = \int_{-\infty}^{\infty} \ud x\, \psi^*\left(x + \frac{s}{2} \right) \psi \left( x - \frac{s}{2} \right),
\end{align}
does not vanish in general. Graphically, this can be seen in Fig.~\ref{introFig}(a) through the overlap of the PSFs. Figure~\ref{introFig}(b) shows, via a complex disk, the possible values of $\Gamma$. Several previously studied cases (with $A = 1$) correspond to highlighted features on this disk: the incoherent, real, and coherent cases correspond to the origin, the real axis, and the circumference of the disk \cite{Tsang2016,Larson2018,Hradil2019}. Part of the goal of this work is to extend these analyses to the rest of the disk; namely, to analyze the quantum FI (for $s$) when the $\Gamma$ between the two separated point sources is described as non-real.

Equation~(\ref{densMat}) indicates that the there are four parameters to be treated in the framework of parameter estimation: $\mathcal{P}=\{A, r,\phi,s\}$. The goal then is to obtain a $4\times4$ quantum FI matrix (QFIM), from which the quantum FI is calculated. This process begins with finding an orthonormal basis that spans $\hat{\rho}$ and its parametric derivatives $\partial_j \hat{\rho}$, where $j \in \mathcal{P}$. Once this basis is obtained, symmetric logarithmic derivative (SLD) matrices for each parameter can be calculated. Finally, the QFIM is readily obtained from the SLD matrices \cite{Liu2019}.

To begin, $\hat{\rho}$ is re-expressed in terms of an orthogonal basis $\{|\psi_+, |\chi_- \rangle\}$, where $|\chi_-\rangle$ is defined as
\begin{align}
    |\chi_- \rangle \triangleq \left(1-d\right)^{-1/2} \left( |\psi_- \rangle - d |\psi_+\rangle \right),
\end{align}
so that $\langle \psi_+ | \chi_- \rangle = 0$. Note that, as is the case throughout, the explicit dependence of $d$ on $s$ is suppressed. In this basis,
\begin{align}
    \hat{\rho} &= \begin{bmatrix} 1 + A d^2 + 2 d r \cos (\phi) & \sqrt{1 - d^2} \left( A d + r e^{\ii \phi} \right)   \\ \sqrt{1 - d^2} \left( A d + r e^{-\ii \phi} \right) & A (1-d^2) \end{bmatrix} \times \nonumber\\
    & [1 +A + 2 d r \cos(\phi)]^{-1}. \label{newdensMat}
\end{align}
With $\hat{\rho}$ written in an orthogonal basis, it is now possible to see details that were initially obscured in the $\{|\psi_+\rangle,|\psi_-\rangle\}$ basis. For example, it is explicit that $\hat{\rho}$ depends on all the parameters in $\mathcal{P}$ (although the $s$-dependence is hidden in $d$) and that $\Tr(\hat{\rho}) = 1$. Note that the unit trace condition reflects the notion that the number of photons arriving at the image plane, for a fixed object plane photon number, depends on the parameters in $\mathcal{P}$. Furthermore, it is straightforward to diagonalize $\hat{\rho}$ in order to obtain eigenkets $|e_i \rangle$, for $i = 1,2$, that span $\hat{\rho}$. However, it is also necessary to also have access to kets that span the matrices $\partial_j \hat{\rho}$ for the computation of the SLD matrices and the QFIM). It is shown in Appendix~\ref{app2} that four orthonormal kets $\mathcal{E} \triangleq \{|e_1\rangle, |e_2 \rangle, |e_3\rangle, |e_4\rangle\}$ are needed to span the $4\times 4$ matrices of $\hat{\rho}$, and $\partial_j \hat{\rho}$. 

With $\mathcal{E}$, it is now possible write down the SLD matrix $\hat{\mathcal{L}}^j$ for each parameter $j \in \mathcal{P}$:
\begin{align}
    \hat{\mathcal{L}}^{j} = \sum_{l,k=1}^4 \frac{2\langle e_k | \partial_j \hat{\rho} |e_l\rangle }{\lambda_k+\lambda_l} |e_k \rangle \langle e_l|,
\end{align}
where the terms for which $\lambda_k + \lambda_l = 0$ are omitted from the sum. The elements of the QFIM can then be calculated as
\begin{align}
    \hat{\mathcal{Q}}_{ij} = \frac{1}{2} \Tr \left[ \left( \hat{\mathcal{L}}^i \hat{\mathcal{L}}^j + \hat{\mathcal{L}}^j \hat{\mathcal{L}}^i \right) \hat{\rho} \right],
\end{align}
where $\hat{\mathcal{Q}}_{ij}$ is the $(i,j)$-th element of the QFIM. Although it is possible to analytically derive the expressions for the SLD matrices and the QFIM, their unruly form does not lead to any obvious insights and it is impractical to state them here. Observe that $\hat{\mathcal{Q}}$ is a $4\times 4$ real and symmetric matrix.

Although the SLD matrices are presented here as an intermediate step in determining the QFIM, they importantly serve the role of determining whether a single measurement of multiple parameters can be optimal with regards to the quantum Cramer-Rao bound determined by the QFIM. Namely, optimal measurements for $i,j \in \mathcal{P}$ can be simultaneously obtained if 
\begin{align}
    \Tr (\hat{\rho} [\hat{\mathcal{L}}^i, \hat{\mathcal{L}}^j ] ) =0, \label{simulcond}
\end{align}
where $[\hat{\mathcal{L}}^i, \hat{\mathcal{L}}^j ]$ is a commutator. 

Finally, upon obtaining the QFIM, it is possible to calculate the quantum FI regarding a measurement done on the parameters in $\mathcal{P}$ under a variety of conditions. First, one must identify the parameters $\mathcal{P_\text{u}} \triangleq \{j_1, \dots, j_m\} \subseteq \mathcal{P}$, where $1\le  m \le |\mathcal{P}|$ that are \textit{unknown} (to be estimated through measurement). Once this is done, one then considers the submatrix $\hat{\mathcal{Q}}(\mathcal{P}_\text{u})$ that contains the rows and columns that correspond to the parameters in $\mathcal{P}_\text{u}$. Note that $\hat{\mathcal{Q}}(\mathcal{P}_\text{u})$ will be a $m \times m$ matrix. The quantum FI $H_{j_i}$, for the parameter $j_i$, with $i=1,\dots,m$, is then obtained as
\begin{align}
    H_{j_i} = \left\{\left[  \hat{\mathcal{Q}}(\mathcal{P_\text{u}})^{-1} \right]_{ii} \right\}^{-1}. \label{qfi}
\end{align}
That is, the quantum FI is the scalar inverse of the $(i,i)$-th element  $\hat{\mathcal{Q}}(\mathcal{P}_\text{u})^{-1}$.

\section{Results} \label{eqint}

Since our analysis involves the potential of calculating multiparameter quantum FI, it is important to understand when this FI is simultaneously achievable. As noted before, this occurs for parameters $i,j \in \mathcal{P}_\text{u}$ when Eq.~(\ref{simulcond}) is satisfied. Of particular interest is the case of $i = s$; that is, when one of the parameters is the separation between the two point sources. The conditions for which Eq.~(\ref{simulcond}) is satisfied for various $j \in \mathcal{P}_\text{u}$ are summarized in Table~\ref{tab:simultable}.

For simplicity, the remainder of our discussion will assume that the intensity PSF takes the form of a Gaussian:
\begin{align}
    \psi(x) = \frac{1}{\sqrt{2\pi} \sigma}\exp \left( -\frac{x^2}{2\sigma^2} \right),
\end{align}
where $\sigma$ is the standard deviation of the PSF. Furthermore, the analysis is predominantly limited to when $A = 1$. This is the case for which most commentary regarding Rayleigh's curse has been generated; indeed, the term "Rayleigh's curse" has its clearest meaning for $A = 1$.  

\begin{table}
    \centering
    \begin{tabular}{c|l}
       $i,j \in \mathcal{P}_\text{u}$  & Sufficient condition for $\Tr (\hat{\rho} [\hat{\mathcal{L}}^i, \hat{\mathcal{L}}^j ] ) =0$  \\ \hline 
        $s,r$ & $A = 1$ or $\text{Im}(\Gamma) = 0$ \\
        $s,\phi$ & $A=1$ or $\text{Re}(\Gamma)=0$\\
        $s,A$ & $\text{Im}(\Gamma) = 0$
    \end{tabular}
    \caption{Sufficient conditions for attainability for simultaneous optimal measurements for estimating $i,j \in \mathcal{P}_\text{u}$ [when Eq.~(\ref{simulcond}) is satisfied]. In particular, note that a sufficient condition for the first two rows, which correspond to the cases of partial knowledge regarding $\Gamma$, is $A = 1$ (equal point source intensities).}
    \label{tab:simultable}
\end{table}

\subsection{Estimating only the separation between two point sources}
\begin{figure}[ht]
    \centering
    \includegraphics[scale=0.4]{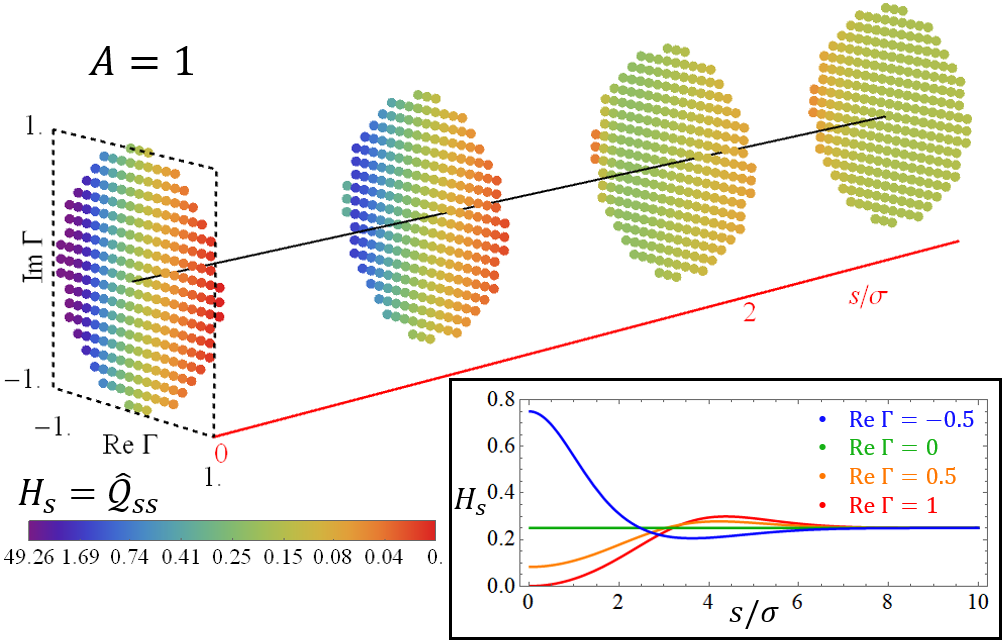}
    \caption{$H_s$, for the case of $A=1$ and $\mathcal{P}_\text{u} = \{s\}$, is shown as a function of complex $\Gamma$ and $s$. The black line that runs longitudinally through the plot corresponds to $\Gamma = 0$ (incoherent point sources). The inset shows several curves of $H_s$ for values of $\Gamma$, highlighting the fact that $H_s$ does not depend on $\text{Im}(\Gamma)$.}
    \label{sHs3D}
\end{figure}

The simplest $H_s$ is derived for the case when $\mathcal{P}_\text{u} = \{s\}$. That is, all the parameters are assumed known aside from $s$. Here, Eq.~(\ref{qfi}) reduces to $H_s = \hat{\mathcal{Q}}_{ss}$. This quantum FI is plotted in Fig.~\ref{sHs3D} as a function of the complex coherence parameter $\Gamma$ (on transverse disks) and the normalized separation $s/\sigma$ (longitudinally). 

The three special cases of Tsang, Larson, and Hradil (which are the incoherent, purely real, and purely coherent cases) are encapsulated in the transverse disks as the origin, the region of $\text{Im}(\Gamma) = 0$, and the circumference of the $\Gamma$ disk, respectively. The inset shows several special cases of $H_s$ that were previously mentioned in by Tsang and Saleh. Furthermore, note that Fig.~\ref{sHs3D} shows that for $\text{Re}(\Gamma) = 0$, $H_s$ is a constant for all $s/\sigma$  with value $H = 1/4$ (olive green color in Fig.~\ref{sHs3D}). This is a generalization of Tsang's result regarding the incoherent case. For all other values of $\Gamma$, $H_s$ asymptotically approaches $H$ as $s/\sigma$ increases. Note that this asymptotic behavior is also true for $H_s$ in the cases, discussed in later sections, where $\mathcal{P}_\text{u}$ contains more unknown parameters than just $s$.

\begin{figure}
    \centering
    \includegraphics[scale=0.45]{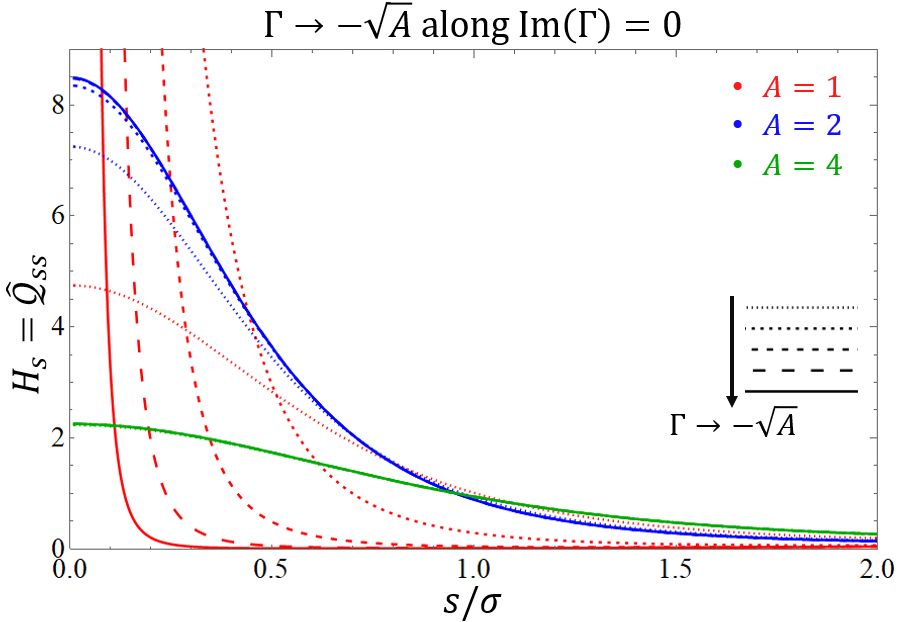}
    \caption{$H_s$ for the case of $\mathcal{P}_\text{u} = \{s\}$, as a function of $s$, is shown for several values of $A$ (the color of the curves) and real $\Gamma$ that approach $\Gamma \rightarrow -\sqrt{A}$ (the dashing of the curves). For $A = 1$, $H_s$ diverges at the infinitesimal point $s = 0$ as $\Gamma \rightarrow -1$. For other values of $A$, however, this anomalous behavior does not exist and the curves of various opacity converge to a finite value at $s = 0$ (the four curves corresponding to $A = 4$ are not distinguishable).}
    \label{anomalousBehavior}
\end{figure}

The transverse disk $s = 0$ depicts the well-known anomalous behavior of $H_s$ as $\Gamma \rightarrow -1$, which is the case where the two point sources become perfectly anti-correlated (the limit is naturally defined along the $\text{Re}(\Gamma)$ axis since $H_s$ is independent of $\text{Im}(\Gamma)$). This behavior, which is further shown in Fig.~\ref{anomalousBehavior}, describes the divergence of $H_s$ as $\Gamma \rightarrow -1$ over an infinitesimally small region of $s/\sigma$. The cause of this anomalous behavior may be traced to the fact that the image of two perfectly anti-correlated point sources is dark when their separation vanishes. Indeed, this is reinforced by Fig.~\ref{anomalousBehavior}, which shows that the anomalous behavior does not exist for other values of $A$. That is, for $A \neq 1$, the value of $H_s$ at $s=0$ and $\Gamma = -\sqrt{A}$ is finite (and relatively large). Therefore, the theory described in Sec.~\ref{theory} indicates that if the intensities of the two sources are known to be even slightly unequal, the value of $H_s$ at $s = 0$, for $\Gamma  = -\sqrt{A}$, is finite and large.

\subsection{Estimating both the separation and another parameter}

The case of $\mathcal{P}_\text{u} = \{j,s\}$, where $j \neq s$, is now considered. The quantum FI is given here by
\begin{align}
    H_s = \hat{\mathcal{Q}}_{ss} - \frac{\hat{\mathcal{Q}}_{js}^2}{\hat{\mathcal{Q}}_{jj}}.
\end{align}

\begin{figure}[ht]
    \centering
    \includegraphics[scale=0.4]{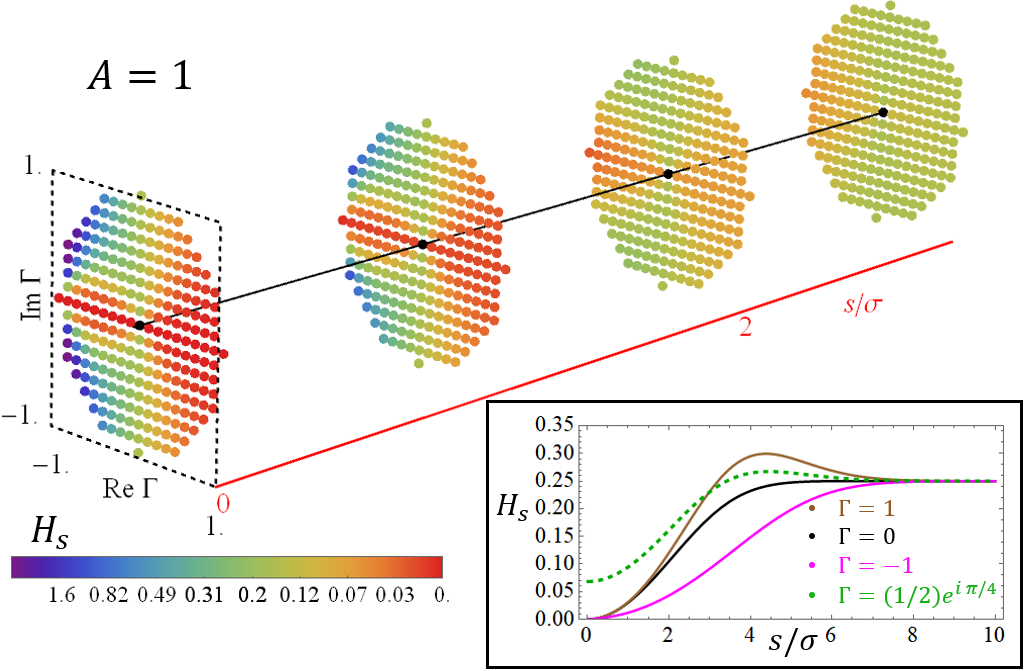}
    \caption{$H_s$, for the case of $A=1$ and $\mathcal{P}_\text{u} = \{r,s\}$, is shown as a function of complex $\Gamma$ and $s/\sigma$. Note that the transverse disk of $s = 0$ indicates that it is possible to violate Rayleigh's curse even if $r$ is unknown along with $s$. The black dots indicate a singular value. The inset shows several curves of $H_s$ for values of $\Gamma$, which shows that $H_s = 0$ on the $s = 0$ disk if $\text{Im}(\Gamma) = 0$. Otherwise, the quantum FI need not vanish.}
    \label{srHs3D}
\end{figure}

The case of $\mathcal{P}_\text{u} = \{r,s\}$ is analyzed first, which includes Larson's analysis in which it was claimed that the lack of knowledge of both the coherence and the separation causes Rayleigh's curse to return ($H_s = 0$ as $s \rightarrow 0$). However, their study was limited to the case of real $\Gamma$, which inherently assumes knowledge regarding $\phi$, the \textit{phase} of the coherence. Therefore, their case actually corresponds to when only $r$ (and not $\phi$) is unknown in addition to $s$. This is precisely the case to be discussed now. Figure~\ref{srHs3D} shows $H_s$ ; note $H_s \neq 0$ over the transverse disk of $s = 0$, which indicates that it is possible to avoid Rayleigh's curse even when $\mathcal{P}_\text{u} =\{r,s\}$. %Although this notion goes against Saleh's claim, 
It should be noted that in Fig.~\ref{srHs3D} through the particular cross-section of $\text{Im}(\Gamma) = 0$, $H_s = 0$ at $s = 0$ (see inset). In other words the case of $\text{Im}(\Gamma) = 0$, where Rayleigh's curse returns, is the only case for which it does. Other values of $\Gamma$ allow for a non-zero $H_s$ at $s = 0$.

\begin{figure}[ht]
    \centering
    \includegraphics[scale=0.4]{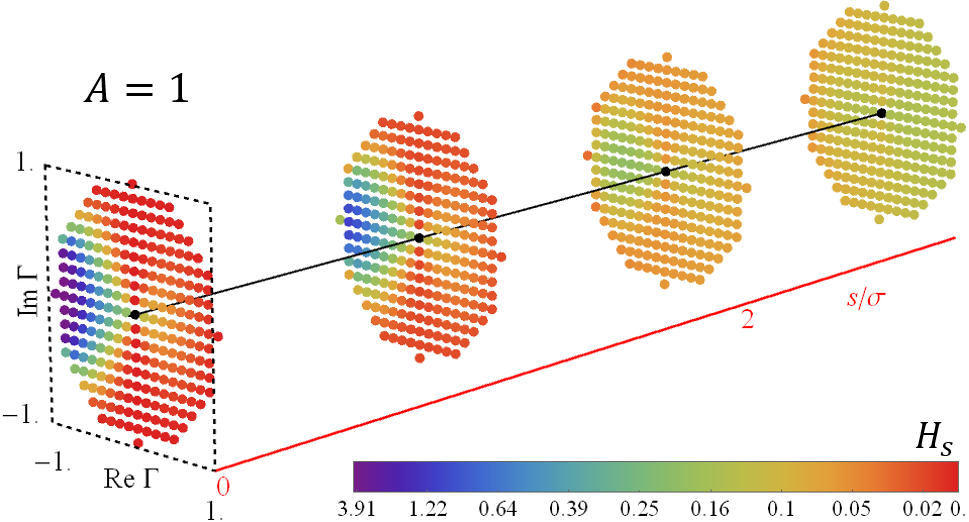}
    \caption{$H_s$, for the case of $A=1$ and $\mathcal{P}_\text{u} = \{\phi,s\}$, is shown as a function of complex $\Gamma$ and $s/\sigma$. Note that the transverse disk of $s = 0$ indicates that it is possible to violate Rayleigh's curse even if $\phi$ is unknown along with $s$. The black dots indicate a singular value. }
    \label{spHs3D}
\end{figure}

One can also consider the case $\mathcal{P}_\text{u} = \{\phi,s\}$. Unlike the preceding case, it is now assumed that $\phi$, the phase of the coherence parameter, is unknown (and $r$ is known). The corresponding $H_s$ is shown in Fig.~\ref{spHs3D}, which indicates, like in the case of $\mathcal{P}_\text{u} = \{r,s\}$, that it is also possible to avoid Rayleigh's curse since $H_s$ does not necessarily vanish over the disk $s = 0$. In fact, it vanishes only when $\text{Re}(\Gamma) = 0$. Note that, in contrast with the case of $\mathcal{P}_\text{u} = \{r,s\}$, this condition involves the zero set of $\text{Re}(\Gamma)$ rather than that of $\text{Im}(\Gamma)$.

\begin{figure}[ht]
    \centering
    \includegraphics[scale=0.55]{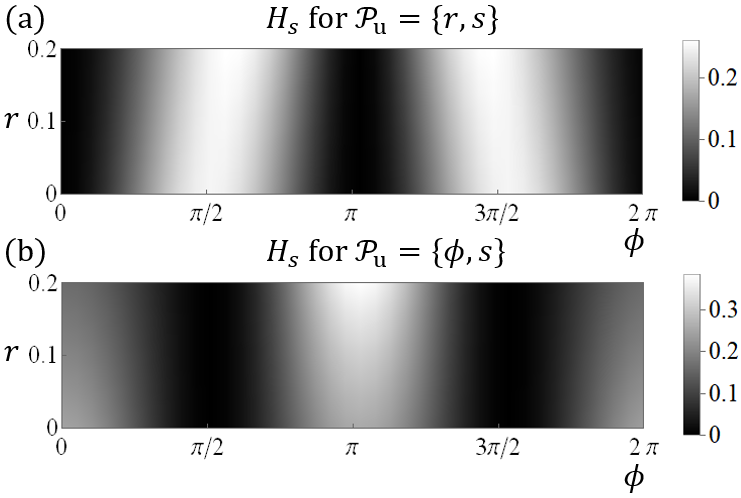}
    \caption{$H_s$, at $s = 0$ is shown, as a function of $r$ and $\phi$, for the cases of $\mathcal{P}_\text{u} = \{r,s\}$ and $\mathcal{P}_\text{u} = \{\phi,s\}$ in (a) and (b), respectively. The singular behavior is reflected by the fact that $H_s$ takes on multiple values along the horizontal line of $r=0$ despite it being a single point in the complex $\Gamma$-disk.}
    \label{singularBehavior}
\end{figure}

Note that $H_s$ is singular at $\Gamma = 0$ for $\mathcal{P}_\text{u} = \{j,s\}$ where $j \in \{r,\phi\}$ (indicated by black dots in Figs.~\ref{srHs3D} and \ref{spHs3D}). This behavior is shown in an alternative manner in Fig.~\ref{singularBehavior}, where the singular nature is represented by the multi-valued nature of $r = 0$. That is, depending on the trajectory one takes (which $\phi$, for instance) in the limit of $\Gamma \rightarrow 0$, $H_s$ approaches a different value. Although Fig.~\ref{singularBehavior} only shows this behavior for $s = 0$, this phenomenon persists as $s$ increases. However, as indicated by Figs.~\ref{srHs3D} and \ref{spHs3D}, the range of multi-values that $H_s$ can take for $r = 0$ collapses to $H = 1/4$ as $s$ increases. Finally, we note that this singular behavior is evident when the analysis includes both $r$ and $\phi$. It is possible to miss this phenomenon if knowledge regarding either $r$ or $\phi$ is assumed to be known in Eq.~(\ref{densMat}). For instance, if one were to assume $\text{Im}(\Gamma) = 0$ (as was done in Saleh's work), then the limit as $\Gamma \rightarrow 0$ is automatically restricted to the vertical trajectories in Fig.~\ref{singularBehavior}(a) that correspond to $\phi \in \{0,\pi\}$. Since $H_s = 0$ for both of these trajectories, the singular behavior is consequently missed. The appearance of this singular behavior is perhaps indicative of some ambiguity regarding $H_s$ when $\Gamma$ is partially known. Presently, it is unclear as to what may determine (theoretically or experimentally) which of the multiple values $H_s$ may take at $\Gamma = 0$.

Only the plots of $H_s$ are shown in Fig.~\ref{srHs3D} and \ref{spHs3D}, since $s$ is arguably the more important parameter to estimate under the framework of this analysis. However, $H_j$, the quantum FI for the other unknown parameter $j$ can also be calculated. The two quantities $H_s$ and $H_j$ then represent the quantum FI for the two unknown parameters $j,s$, keeping in mind that these information bounds can be simultaneously reached (measured) only if the conditions in Table~\ref{tab:simultable} are satisfied. 

Although it is possible to consider the case of $\mathcal{P}_\text{u} = \{A,s\}$, this case is not relevant to the analysis of how partial ignorance of $\Gamma$ affects $H_s$. Therefore, for brevity, this case is not further discussed despite the fact that the theory in Sec.~\ref{theory} fully encapsulates this route of inquiry as well.

\subsection{Estimating the separation and the complex coherence}

\begin{figure}
    \centering
    \includegraphics[scale=0.4]{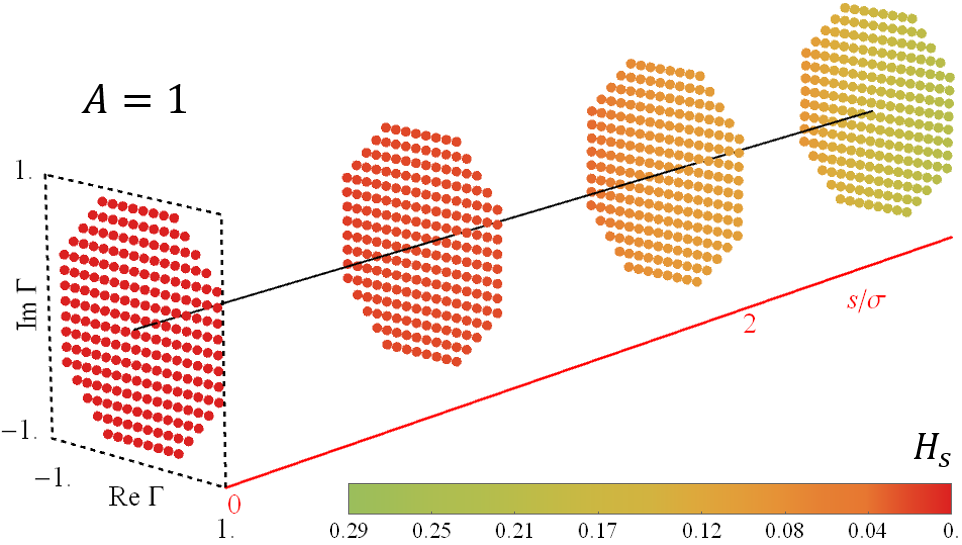}
    \caption{$H_s$, for the case of $A=1$ and $\mathcal{P}_\text{u} = \{r,\phi,s\}$, is shown as a function of complex $\Gamma$ and $s/\sigma$. The black line that runs longitudinally through the plot corresponds to $\Gamma = 0$ (incoherent point sources). $H_s$ is zero over the entire $s=0$ transverse disk; it is not possible to circumvent Rayleigh's curse in this case.}
    \label{srpHs3D}
\end{figure}

The case of $\mathcal{P}_\text{u} = \{r,\phi,s\}$ is now considered. This corresponds to the situation of where, in addition to $s$, both the magnitude and phase of $\Gamma$. With three unknown parameters, $H_s$ is given by the more complicated expression of
\begin{align}
    H_s &= \hat{\mathcal{Q}}_{ss}+ \frac{\hat{\mathcal{Q}}_{rs}^2\hat{\mathcal{Q}}_{\phi\phi}-2\hat{\mathcal{Q}}_{r\phi}\hat{\mathcal{Q}}_{rs}\hat{\mathcal{Q}}_{\phi s}+ \hat{\mathcal{Q}}_{rr} \hat{\mathcal{Q}}_{\phi s}}{\hat{\mathcal{Q}}_{r\phi}^2 - \hat{\mathcal{Q}}_{rr}\hat{\mathcal{Q}}_{\phi\phi}},
\end{align}
and is plotted in Fig.~\ref{srpHs3D}. There, it is apparent that when no information about $\Gamma$ is known, Rayleigh's curse is unavoidable since $H_s$ vanishes over the entire $s=0$ disk. Figure ~\ref{srpHs3D} shows that \textit{complete} ignorance of the coherence parameter $\Gamma$ does indeed lead to a vanishing quantum FI for $s$. However, despite the fact that $H_s$ vanishes here at $s = 0$, the FI is still larger than the classical FI of direct intensity measurements in a comparable scenario for $s > 0$. Therefore, the quantum FI calculations here indicate a possible advantage over conventional imaging methods even when $\Gamma$ is completely unknown. Further discussion is found in the Appendix~\ref{app3}.

Although it is possible to look at other combinations of three unknown parameters (namely those that include $A$ as an unknown), those results are not shown here since the main purpose of this work is to analyze the relationship between the quantum FI for $s$ and how it is affected by the ignorance (partial or full) of $\Gamma$. Finally, it is also possible to consider $\mathcal{P}_\text{u} = \mathcal{P}$; that is, the case where all four parameters are considered unknown. However, it is not possible for $H_s$ to increase from that shown in Fig.~\ref{srpHs3D} since having additional unknowns will only serve to lower $H_s$. Therefore, Rayleigh's curse cannot be avoided when all four parameters in $\mathcal{P}$ are unknown.

\section{Concluding Remarks}

The question of whether the ignorance of coherence, $\Gamma$, causes the resurgence of Rayleigh's curse is shown to be more complicated when its magnitude and phase ($r$ and $\phi$, respectively) are considered to be distinct parameters to be estimated. To properly address this question, a theoretical framework for the multi-parameter estimation of two point sources is introduced, where the set of possibly-unknown parameters, $\mathcal{P} = \{A,r,\phi,s\}$, consists of the intensity ratio, magnitude of coherence, phase of coherence, and the separation between the two sources. 

The simplest case, where only the separation is unknown, corroborated the previously known results regarding the violation of Rayleigh's curse. Namely, it is shown that as long as $\text{Re}(\Gamma) \neq 1$ (for $A = 1$), then $H_s$, the quantum FI for source separation, is non-zero. Unsurprisingly, the anomalous behavior for when the sources are fully anti-correlated persists in the present framework. However, this anomaly disappears when the intensities of the two sources are unequal. 

The main results of this work concern the cases where $\Gamma = r e^{\ii \phi}$ is partially or completely unknown. When $\phi$ is known and $r$ is unknown, $H_s$ is shown to vanish over the $s = 0$ disk for the region $\text{Im}(\Gamma) = 0$. This result agrees with a previous work's assertion that, for real $\Gamma$, $H_s$ vanishes with the ignorance of $r$. However, we have shown that the region defined by $\text{Im}(\Gamma) = 0$ is the only one where $H_s$ vanishes. For all other (complex) values of $\Gamma$, it is evidently possible to break Rayleigh's curse. Similar results are true for when $r$ is known and $\phi$ is unknown. Hence, when $\Gamma$ is partially known within the context of multi-parameter estimation, it is possible to break Rayleigh's curse. However, when $\Gamma$ is completely unknown, $H_s$ vanishes over the $s = 0$ disk. Therefore, complete ignorance of $\Gamma$, unlike partial ignorance (where either $r$ or $\phi$ is known), necessarily causes Rayleigh's curse to reappear. Additional findings include the singular behavior at $\Gamma = 0$ for the cases when $\Gamma$ is only partially known.

The theoretical framework developed in this work allows for a wider scope of analysis than the results presented. This includes the effects of considering unequal intensities ($A \neq 1$), allowing $A$ to be an unknown parameter, and analyzing the quantum FI, $H_j$, for parameters $j \neq s$ that would need to be jointly estimated with the separation $s$. Additional discussions are possible regarding the attainability of optimal joint measurements and possible experimental verification of the results regarding the partial ignorance of $\Gamma$. However, these aforementioned topics, although interesting in their own right, fall outside the scope of this manuscript, which primarily serves to clarify the effects of coherence in two-point separation estimation. Nevertheless, an overview of an experimental design, used in Ref.~\onlinecite{Wadood2020}, that can explore the theoretical results is provided as follows: the measured intensity arising from two partially coherent point sources can be generated by summing the intensities from its two coherent modes via the coherent mode decomposition. These modes can be generated from Gaussian laser light passing through a mode converter, and their intensities measured from a parity sorter to perform a binary SPADE measurement. By adjusting the weights of the intensity summation between the two modes and performing the subsequent maximum likelihood estimation on the measured intensities, it is possible to explore the FI from different portions of the complex coherence disk.

\appendix

\section{Density matrix for two point sources} \label{app1}
The derivation of the density matrix $\hat{\rho}$ follows that of Tsang's work (see Appendix B of Ref.~\onlinecite{Tsang2016}). To summarize, one begins with the coherent state (Sudarshan-Glauber) representation of the density operator 
\begin{align}
    \hat{\rho} = \int \Phi(v) |v\rangle \langle v| \, \ud^{2M} v, \label{SGrep}
\end{align}
where the integral is over the entire complex phase space in which the coherent state is defined and $v = [v_1,\dots, v_M]^\text{T}$ is a column vector of complex field (coherent) amplitudes for $M$ optical space modes on the image plane. That is, $|v\rangle$ is a multimode coherent state with (vector) amplitude $v$. 
\begin{comment}
Assuming that the two sources are thermal, the weight function of the coherent state representation, $\Phi(v)$, is given by
\begin{align}
    \Phi(v) = \frac{1}{\det (\pi \hat{G} )} \exp \left(-v^\dagger \cdot \hat{G} \cdot v \right), \label{thermalSource}
\end{align}
where $\hat{G}$ is the image plane mutual (spatial) coherence matrix. Note that Eq.~(\ref{thermalSource}) is a particularly simple weight function as it is real and positive. These properties (which are not in general true for other types of sources) imply that the proper quantum optical treatment of thermal sources is equivalent to its corresponding classical one. 
\end{comment}
The probability of having $j$, total photons, $p_n$, is then given by $p_j = \Tr \left(\hat{\rho}|\{j\}\rangle\langle \{j\} | \right)$, where $|\{j\}\rangle$ is the $j$ photon multimode state.

Several reasonable assumptions are now considered. The average number of photons arriving at the image plane during the coherence time of the source, $\epsilon$, is considered to be much smaller than 1. That is,
\begin{align}
    \epsilon \triangleq \langle \hat{n} \rangle =  \sum_{m = 0}^M \Tr \left( \hat{\rho} \hat{n}_m \right) \ll 1,
\end{align}
where $\hat{n}$ is the multimode photon number operator, $\hat{n}_m$ is the photon number operator for the $m$-th mode and $\langle \cdot \rangle$ denotes an expectation value with respect to the representation in Eq.~(\ref{SGrep}). One can rewrite $\epsilon$, using $\hat{n} = \hat{a}^\dagger \hat{a}$ as
\begin{align}
    \epsilon  = \langle  \hat{a}^\dagger \hat{a} \rangle = \langle |v|^2 \rangle  = \int  \Phi(v) |v|^2 \, \ud^{2M}v \ll 1, \label{epsilon}
\end{align}
where $\hat{a}^\dagger$ and $\hat{a}$ are the multimode creation and annihilation operators. In the last step of Eq.~(\ref{epsilon}), the operators are replaced with their corresponding eigenvalues since Eq.~(\ref{SGrep}) is a coherent state representation. 

The condition in Eq.~(\ref{epsilon}) implies several more useful simplifications. In particular, note that $p_j$ is given by
\begin{align}
    p_j &= \int \Phi(v) |\langle v | \{j\} \rangle|^2\, \ud^{2M}v = \frac{1}{j!}\left\langle \exp\left(-|v|^2\right) |v|^{2j}       \right\rangle, 
\end{align}
which, upon Taylor expansion of the exponential, gives
\begin{align}
    p_j &=  \frac{1}{j!} \sum_{k=0}^\infty \frac{(-1)^k}{k!} \left\langle  |v|^{2(k+j)}  \right\rangle \nonumber\\
    &= \frac{1}{j!} \left[ \langle |v|^{2j}\rangle - \langle |v|^{2(1+j)} \rangle \right] + O\left(\epsilon^{j+2} \right).
\end{align}
Since $\epsilon\ll 1$, only terms up to linear order in $\epsilon$ are significant. This can only happen for $j = 0$ and $j = 1$. That is, we arrive at
\begin{align}
    p_0 &= \langle 1 \rangle - \langle |v|^2 \rangle + O(\epsilon^2) = 1 - \epsilon + O(\epsilon^2), \label{p0} \\
    p_1 &= \langle |v|^2 \rangle - \langle |v|^4 \rangle + O(\epsilon^3) = \epsilon + O(\epsilon^2),\label{p1}\\
    p_{j\ge 2} &= O(\epsilon^2). \label{p2}
\end{align}
Note that we used the fact that $\langle 1 \rangle = 1$, which results from $\Tr(\hat{\rho}) = 1$. Equations~(\ref{p0}) - (\ref{p2}) indicate that only that multi-photon events are insignificant when compared to the zero-photon and one-photon events. Moreover, since the zero-photon event (vacuum state) provides no information regarding measurements, it is actually only the one-photon event, corresponding to $p_1$, that should be examined. 

This particular event corresponds to an element of $\hat{\rho}$ in its Fock (number) state representation. That is, one can consider 
\begin{align}
    \hat{\rho} = \sum_{n=0}^\infty \sum_{m=0}^\infty \langle \{m \}| \hat{\rho}|\{n\}\rangle |\{m\}\rangle \langle \{n\}|, \label{rhoinFock}
\end{align}
where the elements $\langle \{m\}|\hat{\rho}|\{n\}\rangle$ can be found through Eq.~(\ref{SGrep}). Note that $\hat{\rho}$ is not necessarily diagonal when represented in the multi-mode Fock basis; nevertheless the preceding discussion regarding the sole significance of the one-photon event allows us to approximate, using $p_1 \approx \epsilon$
\begin{align}
    \hat{\rho} \approx \epsilon \left( \frac{1}{\epsilon} |\{1\}\rangle \langle \{1\}| \right) = \epsilon \hat{\rho}_1,
\end{align}
where $\hat{\rho}_1$ is the one-photon multi-mode Fock state. Note that, in order for $\hat{\rho}$ to maintain unit trace, an additional $\epsilon^{-1}$ factor had to be introduced to the definition of $\hat{\rho}_1$ in relation to $|\{1\}\rangle\langle \{1\}|$. 

At this point, through the choice of only considering one-photon events, we shift our focus from the entire $\hat{\rho}$ to just $\hat{\rho}_1$. That is, the density matrix we are after is $\hat{\rho}_1 = \hat{\rho}/\epsilon$. Although subtle, this choice is important due to the normalization of density matrices, which is detailed later. Note that $\hat{\rho}_1$ can be decomposed into a sum of single-mode one-photon states by considering the one-photon basis kets $|1_m\rangle$, where $m=1,\dots,M$. That is,
\begin{align}
    \hat{\rho}_1 \approx \frac{1}{\epsilon}\sum_{j=1}^M\sum_{k=1}^M \langle 1_j |\hat{\rho}|1_k\rangle |1_j\rangle \langle 1_k|,
\end{align}
where 
\begin{align}
    \langle 1_j|\hat{\rho}|1_k\rangle  &= \int \Phi(v) \langle 1_j | v \rangle \langle v| 1_k \rangle \, \ud^{2M}v \nonumber\\
    &= \int \Phi(v) \exp[-(|v_j|^2+|v_k|^2)/2)] v_j v^*_k \, \ud^{2M}v\nonumber\\
    &\approx \int \Phi(v) v_j v^*_k \, \ud^{2M}v, \label{matmat}
\end{align}
where, in the final line, only the zeroth order Taylor series term for the exponential was retained [in accordance to terms of $O(\epsilon^2)$ being insignificant]. Note then, that Eq.~(\ref{matmat}) is, by definition, the $(j,k)$-th element of the image-plane mutual coherence matrix $\hat{G}$. With this identification,
\begin{align}
    \hat{\rho}_1 \approx \frac{1}{\epsilon} \sum_{j=1}^M\sum_{k=1}^M \hat{G}_{jk} |1_j\rangle \langle 1_k|, \label{newRhoMat}
\end{align}
and now it remains to determine $\hat{G}$. For an imaging system, the image-plane mutual coherence matrix is related to the object-plane mutual coherence, $\hat{G}_0$, through
\begin{align}
    \hat{G} = \hat{S} \hat{G}_0 \hat{S}^\dagger, \label{coherenceprop}
\end{align}
where $\hat{S}$ is the system's field scattering matrix [often not unitary and closely related to the well-known point spread function (PSF) in classical optics]. Assuming that the imaging system operates under paraxial conditions, it is possible to use localized wave-packet modes as a basis. In other words, the modes $|1_j\rangle$ can be replaced with $|x_j\rangle = |1_j\rangle/\sqrt{\ud x}$, with $\ud x$ the spacing in the position space. These are discrete \textit{position} kets whose position eigenvalues are given by $x_j = x_0 + j \ud x$, where $x_0$ is an arbitrary origin. At this point, we specialize to the case where the object plane consists of two point sources located at $w_+$ and $w_-$. The object-plane mutual coherence is given by
\begin{align}
    \left(\hat{G}_0\right)_{uv} &= \epsilon_0 [ \delta_{uv} \left( \delta_{uw_+}+ A \delta_{uw_-} \right)  \nonumber\\
    &+ \Gamma \delta_{uw_+}\delta_{vw_-} + \Gamma^* \delta_{uw_-}\delta_{vw_+} ]
\end{align}
where $A$ is the (relative to $w_+$) intensity of the point source at $w_-$, $\Gamma$ is the unnormalized coherence parameter between the two point sources, and $\delta_{nm}$ is the Kronecker delta symbol. Using Eq.~(\ref{coherenceprop}), we find that
\begin{align}
    \hat{G}_{jk} &= \epsilon_0 \bigg( \hat{S}_{jw_+}\hat{S}^*_{kw_+} + A \hat{S}_{jw_-}\hat{S}^*_{kw_-} \nonumber\\
    &+ \Gamma \hat{S}_{j w_+} \hat{S}^*_{k w_-} + \Gamma^* \hat{S}_{j w_-} \hat{S}^*_{kw_+}\bigg).
\end{align}

We return now to $\epsilon$, the average number of photons within a coherence time. It is related to the the scattering matrix elements and the coherence parameter $\Gamma$ through
\begin{align}
    \epsilon = \Tr\left( \sum_{j=1}^M\sum_{k=1}^M \hat{G}_{jk} |1_j\rangle \langle 1_k| \right) \epsilon_0 \eta,
\end{align}
where $\eta \triangleq \sum_{j=1}^M |S_{jw_s}|^2$, with $s \in \{+,-\}$, is the quantum efficiency (assumed to be equal for both point sources). Note that this value of $\epsilon$ ensures that $\Tr(\hat{\rho}_1) = 1$ and explicitly demonstrates that the number of photons arriving at the image plane, for a fixed object plane photon number, depends on the possibly-unknown parameters of $\{A, r, \phi, s\}$. In order to express $\hat{\rho}$ in the familiar basis of two shifted PSFs, we consider the following relations:
\begin{align}
    |\psi_s\rangle \triangleq \sum_{j=1}^M \frac{S_{j,w_s}}{\sqrt{\eta}}|x_j \rangle \sqrt{\ud x} \quad \text{and} \quad \psi_s (x_j) = \frac{S_{j,w_s}}{\sqrt{\ud x}},
\end{align}
one can then take the continuous-space limit of $\ud x \rightarrow 0$ (and hence $M \rightarrow \infty$) to arrive at 
\begin{align}
    \hat{\rho}_1 = \frac{1}{(1+A)[1+d^2(s)] + 4 d(s) \text{Re}(\Gamma)}\begin{bmatrix} 1 & \Gamma  \\ \Gamma^* & A \end{bmatrix}_{\text{no}}, \label{densityMatrix}
\end{align}
which is in the not-orthogonal basis of $\{|\psi_+\rangle, |\psi_-\rangle\}$, as desired. Note that in the main body, for simplicity, this $\hat{\rho}_1$ is labeled as just $\hat{\rho}$.

\section{Obtaining an orthonormal basis} \label{app2}
The explicit process for obtaining a set of orthonormal vectors that spans the space of $\hat{\rho}$, given by (in the orthogonal $\{|\psi_+\rangle, |\chi_-\rangle\}$ basis)
\begin{align}
    \hat{\rho} &= \begin{bmatrix} 1 + A d^2 + 2 d r \cos (\phi) & \sqrt{1 - d^2} \left( A d + r e^{\ii \phi} \right)   \\ \sqrt{1 - d^2} \left( A d + r e^{-\ii \phi} \right) & A (1-d^2) \end{bmatrix} \times \nonumber\\
    & [1 +A + 2 d r \cos(\phi)]^{-1},
\end{align}
and $\partial_j \hat{\rho}$, where $j \in \mathcal{P} = \{A,r,\phi,s\}$, is shown here. First, it is straightforward to diagonalize $\hat{\rho}$ in order to obtain two (normalized) eigenvectors $|e_1\rangle$ and $|e_2\rangle$, which automatically span $\hat{\rho}$. Of course, once diagonalized, $\hat{\rho}$ can be expressed simply as
\begin{align}
    \hat{\rho} &= \sum_{i=1}^2 \lambda_i |e_i \rangle \langle e_i|,
\end{align}
where $\lambda_i$ are the eigenvalues that correspond to $|e_i\rangle$ for $i = 1,2$.

The next step is to find the eigenvectors to $\partial_j \hat{\rho}$, where
\begin{align}
    \partial_j \hat{\rho} &= \sum_{i=1}^2 \left[ (\partial_j \lambda_i) |e_i\rangle \langle e_i | +  \lambda_i \left( |f_i^j \rangle \langle e_i |  + |e_i\rangle \langle f_i^j |\right) \right], \label{partialRho}
\end{align}
where
\begin{align}
    |f^j_i \rangle &\triangleq \partial_j |e_i \rangle. \label{definitionf}
\end{align}
First, the case of $j \neq s$, which turns out to be the simpler case, is analyzed. Given Eq.~(\ref{partialRho}), it is desirable to rewrite the expression of $\partial_j \hat{\rho}$ in terms of the $\hat{\rho}$-spanning eigenvectors $|e_i\rangle$. In order to do this, we first note that these eigenvectors can be expressed in terms of the non-orthogonal basis kets $|\psi_\pm \rangle$ through 
\begin{align}
    |e_i\rangle = F_{ik} |\psi_k \rangle, \label{transferMat}
\end{align}
where $k = 1$ and $k=2$ correspond to $+$ and $-$, respectively. The transformation matrix elements $F_{ik}$ can be easily obtained in the diagonalization process of $\hat{\rho}$ and relating the kets $\{|\psi_+\rangle, |\chi_-\rangle\}$ back to $\{|\psi_+ \rangle, |\psi_- \rangle\}$. Using Eqs.~(\ref{definitionf}) and (\ref{transferMat}), we find that the second term in Eq.~(\ref{partialRho}) can be expressed as
\begin{align}
    \lambda_i  |f_i^j \rangle \langle e_i | + \text{H.C.} &= \lambda_i B^j_{il} |e_l \rangle \langle e_i | + \text{H.C.},
\end{align}
where $B^j_{il} \triangleq (\partial_j F_{ik}) (F^{-1})_{kl}$. It turns out that, for $j \neq s$, the diagonal terms $B^j_{11}$ and $B^j_{22}$ are purely imaginary and therefore do not contribute further. Using this fact, the matrix $\partial_j \hat{\rho}$, for $j \neq s$, can be written in the $\{|e_1\rangle,|e_2\rangle\}$ basis (indicated by a subscript $e$) as 
\begin{align}
    \partial_j \hat{\rho} &= \begin{bmatrix} \partial_j \lambda_1 & \lambda_1 \left(B^j_{12}\right)^* + \lambda_2 B^j_{21} \\ \lambda_1 B^j_{12} + \lambda_2 \left(B^j_{21}\right)^* & \partial_j \lambda_2   \end{bmatrix}_e, \label{partialRhojnots}
\end{align}
where the Hermiticity of $\partial_j\hat{\rho}$ is readily apparent (recall that $\lambda_1$ and $\lambda_2$ are real). Note that the $\partial_j \hat{\rho}$ remains spanned by $\{|e_1\rangle,|e_2\rangle\}$ for $j \neq s$. This was expected because the original basis states $\{|\psi_+\rangle, |\psi_- \rangle\}$ do not depend on $j \neq s$. 

We now look at the remaining case of $j = s$, which, as noted in the discussion after Eq.~(\ref{partialRhojnots}), is complicated by the fact that the original basis states themselves depend on $s$ through the point spread function $\psi$:
\begin{align}
    |\psi_\pm \rangle &\triangleq \int_{-\infty}^{\infty} \ud x \, \psi\left( x \pm \frac{s}{2} \right) |x \rangle. 
\end{align}
Because of this, $\partial_s \hat{\rho}$ is insufficiently spanned by $\{|e_1\rangle, |e_2\rangle\}$ and requires additional kets. Evidently, these additional kets are $\{|f^s_1\rangle, |f^s_2\rangle\}$. For the case of $r e^{\ii \phi} \in \mathbb{R}$, both of these additional kets are automatically (through the definition in Eq.~(\ref{definitionf}) orthogonal to $\{|e_1\rangle, |e_2\rangle\}$ and the construction of an orthonormal basis that spans $\hat{\rho}$ and $\partial_j \hat{\rho}$ is completed through the normalization of $\{|f^s_1\rangle, |f^s_2\rangle\}$. However, in the more general setting explored in this work, this simplifying fact is not true. Nevertheless, it is still relatively straightforward (through the Gram-Schmidt process) to compute the additional basis kets, $|e_3\rangle$ and $|e_4\rangle$, needed to span $\partial_s \hat{\rho}$. That is, we take
\begin{align}
    |e_3\rangle &\triangleq N_3 \left( |f_1^s\rangle - \sum_{p=1}^2 \langle e_p| f_1^s\rangle |e_p\rangle \right), \label{e3expand}\\
    |e_4\rangle &\triangleq N_4 \left( |f_2^s\rangle - \sum_{p=1}^3 \langle e_p| f_2^s\rangle |e_p\rangle \right), \label{e4expand}
\end{align}
where $N_3$ and $N_4$ are normalization constants to ensure $\langle e_3 | e_3\rangle = \langle e_4 | e_4 \rangle = 1$. Equations~(\ref{e3expand}) and (\ref{e4expand}) can be used to replace $|f^s_i \rangle$ in Eq.~(\ref{partialRho}) with $|e_3\rangle$ and $|e_4\rangle$ in order to obtain an expression of $\partial_s \hat{\rho}$ in the orthonormal basis $\mathcal{E} \triangleq \{|e_1\rangle, |e_2\rangle, |e_3\rangle, |e_4\rangle\}$.

The set $\mathcal{E}$ forms an orthonormal basis for $\hat{\rho}$ and $\partial_j \hat{\rho}$, which are now to be extended into $4\times4$ matrices. Aside from $j = s$, only the top-left $2\times2$ submatrix of each is non-zero because $|e_3\rangle$ and $|e_4\rangle$ were needed only to span $\partial_s \hat{\rho}$.

\section{Comparison to direct intensity measurements} \label{app3}
For a real PSF, $\psi$, the photon probability density at the image plane from two partially coherent, equal intensity, point sources is given by
\begin{align}
    P_\text{DI} (x) = \frac{ \psi_+^2(x) + \psi_-^2(x) + 2 \text{Re}(\Gamma) \psi_+(x)\psi_-(x)}{2[1+\text{Re}(\Gamma)d(s)]},
\end{align}
where $x$ is the image plane coordinate, $\Gamma = r e^{\ii \phi}$ is the degree of coherence, and $d(s)$ is the overlap between $\psi_+(x)$ and $\psi_-(x)$. 

In order to calculate the classical Fisher information (FI) with respect to the separation $s$, one must construct a classical FI matrix, whose elements are given by
\begin{align}
    \mathcal{F}_{ij}(\mathcal{P}_\text{u}) = \int_{-\infty}^\infty \frac{1}{P_\text{DI}(x)} [\partial_i P_\text{DI}(x) ][\partial_j P_\text{DI}(x)] \, \ud x,
\end{align}
where $i,j \in \mathcal{P}_\text{u}$, the set of unknown parameters. The classical FI for the parameter $s$ is then given by
\begin{align}
    F_{ss} = \{ [\mathcal{F}(\mathcal{P}_\text{u})^{-1}]_{ss}  \}^{-1}
\end{align}

\begin{figure}
    \centering
    \includegraphics[scale=0.4]{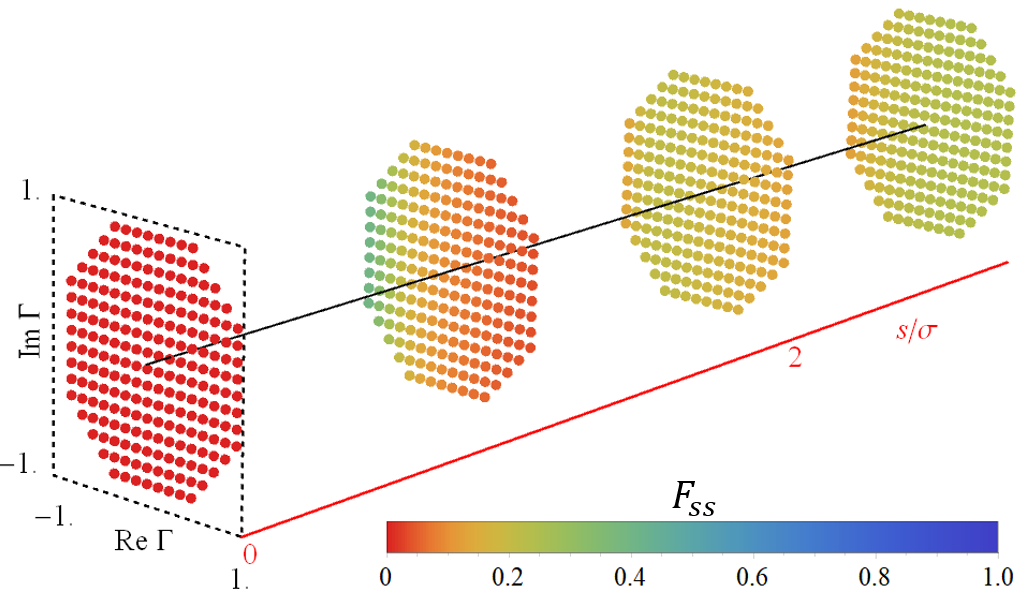}
    \caption{The classical FI, $F_{ss}$, for direct intensity measurement is shown as a function of $s$ and complex $\Gamma$. Here, the set of unknown parameters is $\mathcal{P}_\text{u} = \{s\}$. Note that $\mathcal{F}_{ss} = 0$ over the $s = 0$ transverse disk; this is an illustration of the traditional Rayleigh's curse.  }
    \label{traditionalRC}
\end{figure}

When $\mathcal{P}_\text{u} = \{s\}$, $F_{ss}$ reduces to
\begin{align}
    \mathcal{F}_{ss} = \int_{-\infty}^\infty \frac{1}{P_\text{DI}(x)} [\partial_s P_\text{DI}(x)]^2\, \ud x.
\end{align}
A plot of $\mathcal{F}_{ss}$ is shown in Fig.~\ref{traditionalRC}. Note that $\mathcal{F}_{ss} = 0$ over the $s = 0$ transverse disk, which indicates the traditional Rayleigh's curse of direct intensity measurements.

\begin{figure}[h]
    \centering
    \includegraphics[scale=0.4]{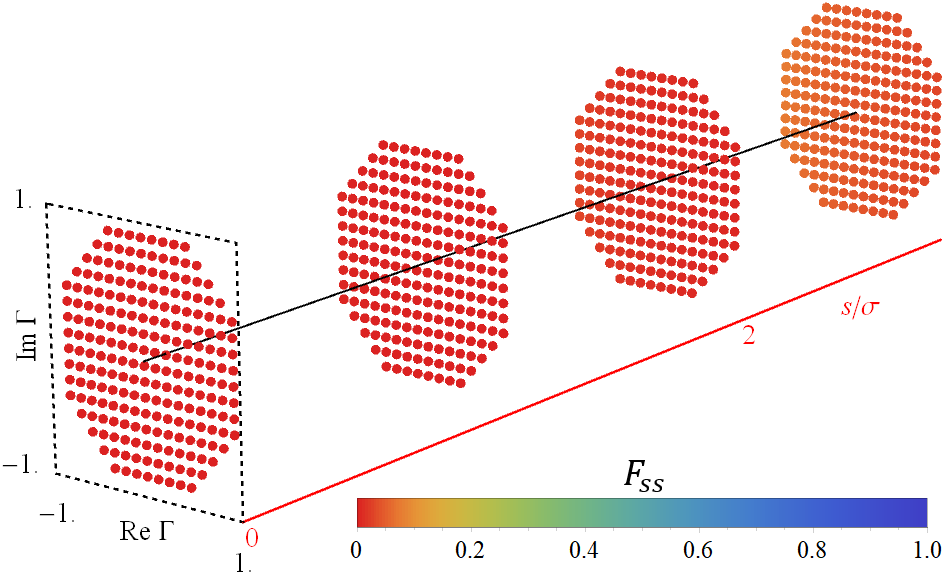}
    \caption{The classical FI, $F_{ss}$, for direct intensity measurement is shown as a function of $s$ and complex $\Gamma$. Here, the set of unknown parameters is $\mathcal{P}_\text{u}=\{r,s\}$. }
    \label{directintensity}
\end{figure}

One can also consider scenarios where there are additional unknown parameters. For instance, consider $\mathcal{P}_\text{u} = \{r,s\}$, which corresponds to the case when the magnitude of the coherence parameter is unknown in additon to the separation. The classical FI $F_{ss}$ for this case is shown in Fig.~\ref{directintensity}. Evidently, the inclusion of an additional unknown parameter drastically lowers the classical FI for direct intensity measurements. In particular, we note that $F_{ss}$ is smaller than the quantum FI for the case of $\mathcal{P}_\text{u} = \{r,\phi,s\}$, which is shown in Fig.~7 of the primary manuscript. This indicates that, even though both situations have vanishing FI over the $s = 0$ transverse disk, the quantum FI calculations still suggest a possible advantage in terms of how the information scales with $s$.

\newpage
\section*{Funding Information}
Defense Advanced Research Projects Agency (D19AP00042)

\section*{Acknowledgments}
The authors thank Andrew N. Jordan for useful discussions.

\section*{Disclosures}
The authors declare no conflicts of interest.

\bibliography{Main}

\end{document}